\documentclass[journal]{IEEEtran}
\usepackage{amsmath,amssymb,amsfonts, bbm}
\usepackage{algorithmic}
\usepackage{graphicx}
\usepackage{textcomp}
\def\BibTeX{{\rm B\kern-.05em{\sc i\kern-.025em b}\kern-.08em T\kern-.1667em\lower.7ex\hbox{E}\kern-.125emX}}
\usepackage{subfigure}
\usepackage{float}
\usepackage{graphicx}
\usepackage{stfloats}
\usepackage{multirow}
\usepackage{tabularx}
\usepackage{amssymb}
\usepackage{booktabs}
\usepackage{hyperref}
\usepackage[table]{xcolor}
\usepackage{tikz}
\usepackage[marginal]{footmisc}
\DeclareMathOperator*{\argmin}{argmin}
\usepackage[switch]{lineno}
\usepackage{bbding}
\usepackage{adjustbox}
\def\ie{\emph{i.e.}}

\def\etal{{\em et al.}}

\definecolor{c1}{HTML}{ffe6e6}
\definecolor{c2}{HTML}{e6e6ff}

\begin{document}
\title{MMR-Mamba: Multi-Modal \\ MRI Reconstruction with Mamba and \\ Spatial-Frequency Information Fusion}

\author{
        Jing Zou,~
        Lanqing Liu,~
        Qi Chen,~
        Shujun Wang,~\IEEEmembership{Member,~IEEE,}\\
        Zhanli Hu,
        Xiaohan Xing,
        and Jing Qin,~\IEEEmembership{Member,~IEEE}\vspace{-15pt}
\thanks{Jing Zou, Lanqing Liu and Jing Qin are with Center for Smart Health, The Hong Kong Polytechnic University, Hong Kong SAR, China (email: zoujing.zou, lanqing.liu, harry.qin@polyu.edu.hk).}
% %
\thanks{Qi Chen is with the Department of Electronic Engineering and Information Science, University of Science and Technology of China, Anhui, China (e-mail: qic@mail.ustc.edu.cn).}
\thanks{Shujun Wang is with the Department of Biomedical
Engineering, The Hong Kong Polytechnic University, Hong Kong
SAR, China (email: shu-jun.wang@polyu.edu.hk).}
\thanks{Zhanli Hu is with Shenzhen Institute of Advanced Technology, Chinese Academy of Sciences, Shenzhen, China (email: zl.hu@siat.ac.cn).}
\thanks{Xiaohan Xing is with the Department of
Radiation Oncology, Stanford University, Stanford, CA 94305 USA (email: xhxing@stanford.edu).}
% %
\thanks{Corresponding author: Xiaohan Xing (xhxing@stanford.edu).}
} 
\maketitle
\begin{abstract}
Multi-modal MRI offers valuable complementary information for diagnosis and treatment; however,  its utility is limited by prolonged scanning times. To accelerate the acquisition process, a practical approach is to reconstruct images of the target modality, which requires longer scanning times, from under-sampled k-space data using the fully-sampled reference modality with shorter scanning times as guidance.
The primary challenge of this task is comprehensively and efficiently integrating complementary information from different modalities to achieve high-quality reconstruction. 
Existing methods struggle with this: 1) convolution-based models fail to capture long-range dependencies; 2) transformer-based models, while excelling in global feature modeling, struggle with quadratic computational complexity.
To address this dilemma, we propose \textit{MMR-Mamba}, a novel framework that thoroughly and efficiently integrates multi-modal features for MRI reconstruction,  leveraging Mamba's capability to capture long-range dependencies with linear computational complexity while exploiting global properties of the Fourier domain.
Specifically, we first design a \textit{Target modality-guided Cross Mamba} (TCM) module in the spatial domain, which maximally restores the target modality information by selectively incorporating relevant information from the reference modality.
Then, we introduce a \textit{Selective Frequency Fusion} (SFF) module to efficiently integrate global information in the Fourier domain and recover high-frequency signals for the reconstruction of structural details. 
Furthermore, we devise an \textit{Adaptive Spatial-Frequency Fusion} (ASFF) module, which mutually enhances the spatial and frequency domains by supplementing less informative channels from one domain with corresponding channels from the other.
Extensive experiments on the BraTS and fastMRI knee datasets demonstrate the superiority of our MMR-Mamba over state-of-the-art reconstruction methods.
\end{abstract}

\section{Introduction}
\label{Intro}

\IEEEPARstart{M}{agnetic} resonance imaging (MRI) is an essential clinical imaging technology due to its non-invasive, free-of-radiation characteristics and its capability to provide high-resolution morphological information with varying modalities \cite{stoja2021improving}.
In clinical practice, multi-modal MR images with complementary information are simultaneously acquired to enable more accurate disease diagnosis and treatment planning\cite{feng2022multi}.
For example, in brain MR imaging, T1 weighted images (T1WIs) provide detailed anatomical structure information, while T2 weighted images (T2WIs) are useful for detecting edema, inflammation, and fluid-filled structures \cite{menze2014multimodal}. Similarly, in knee imaging, proton density weighted images (PDWIs) reveal structural information while fat-suppressed proton density weighted images (FS-PDWIs) can suppress fat signals and highlight cartilage ligaments \cite{chen2015accuracy}. 
However, MR imaging is inherently time-consuming due to sequential data acquisition in k-space, 
%due to the intrinsic physics of MR imaging systems \cite{plenge2012super}, acquiring fully sampled k-space data can take tens of minutes,  r
which results in patient discomfort and increased operational costs \cite{plenge2012super}.
Therefore, accelerating MR imaging, particularly through the reconstruction of high-quality MR images from under-sampled k-space data, is highly demanded in clinical practice \cite{guo2023reconformer}.

%
%Typically, T1WIs are acquired more quickly due to shorter repetition time (TR) and echo time (TE) requirements, whereas T2WIs acquisition requires longer TR times, resulting in slower imaging speeds \cite{xiang2018deep}. 
%
%Similarly, FS-PDWIs necessitate longer scan times than PDWIs. 
%
Previous studies \cite{bilgic2011multi,song2019coupled,lai2017sparse,xiang2018deep,sun2019deep,lyu2023region,li2022transformer} have demonstrated that it is a promising way to  leverage readily obtainable modalities (\ie~reference modalities such as T1WIs or PDWIs) as supplementary guidance for the reconstruction of target modalities (e.g., T2WIs or FS-PDWIs) with slower imaging speeds, known as multi-modal MRI reconstruction.
The primary challenge of this reconstruction task is to comprehensively explore long-range dependencies within each modality and effectively leverage complementary information from reference modalities. %, as similar structural features are distributed across different regions within each modality, while complementary information is provided by different modalities. 
%For instance, compressed sensing (CS), Bayesian learning, dictionary learning, and graph representation theory have been utilized to accelerate multi-modal MR imaging \cite{bilgic2011multi,song2019coupled,lai2017sparse}. More recently, deep learning has significantly advanced MRI reconstruction due to their powerful feature representation capabilities  \cite{xiang2018deep,sun2019deep,song2019coupled,lyu2023region,li2022transformer}. 
%
Early methods employ convolutional neural networks (CNNs) to integrate multi-modal information \cite{xiang2018deep,lyu2020multi,xuan2022multimodal}, but they typically demonstrate local sensitivity and a lack of long-range dependency, thereby limiting their ability to integrate correlated features from both modalities for faithful MRI reconstruction.
%\textcolor{red}{Additionally, the complementary information between multi-modal images is not sufficiently harnessed, as they either directly concatenate images in the input layer \cite{xiang2018deep} or stack features in deep layers \cite{lyu2020multi}.}
%
In contrast, Transformer-based models \cite{feng2022multi,li2023multi,huang2023accurate}, distinguished by their large receptive fields and global sensitivity, often surpass CNNs in capturing extensive contextual information. However, these models are burdened by substantial computational overhead due to the quadratic growth of resources with respect to sequence length. 
Therefore, developing an algorithm that can comprehensively explore long-range dependencies and integrate complementary information from different modalities without significant computational overhead is crucial.

Recently, the improved structured state-space sequence model with a selective scanning mechanism, Mamba \cite{gu2023mamba}, has emerged as a compelling alternative to Transformer, due to its ability to model long-range sequence relationships with linear complexity.
Mamba has shown superior performance compared to Transformers in tasks involving long-term dependency modeling, such as natural language processing \cite{gupta2022diagonal,qin2023toeplitz} and medical image segmentation \cite{xing2024segmamba,ma2024u}.
% and classification \cite{yue2024medmamba,yang2024cmvim}.
%
Investigating Mamba's potential for long-range dependency modeling and complementary information fusion in multi-modal MRI reconstruction is highly promising.
On the other hand, each component in the frequency domain represents a combination of all the pixel values in the spatial domain, meaning that frequency features capture the overall patterns and structures, providing a global view of the entire image. 
Meanwhile, existing studies indicate that Fourier features are beneficial for recovering high-frequency signals that are crucial for addressing image degradation\cite{tancik2020fourier}.
Therefore, comprehensive and efficient global feature integration across different modalities can be achieved by performing feature fusion in the frequency domain.
Thus, Mamba block and frequency domain offer two promising solutions for comprehensive and efficient fusion of multi-modal information.

Motivated by the above analysis, we propose a novel framework, MMR-Mamba, for multi-modal MRI reconstruction. 
Built upon the Mamba architecture, our MMR-Mamba jointly explores complementary information fusion in the spatial and frequency domains, implemented by the \textit{Target modality-guided Cross Mamba} (TCM) module and \textit{Selective Frequency Fusion} (SFF) module, respectively. 
Additionally, we introduce an \textit{Adaptive Spatial-Frequency Fusion} (ASFF) module to mutually enhance the features from these two domains.
Specifically, Mamba blocks are employed to extract features from each modality.
Then we design  TCM for spatial domain information fusion, where correlated features from the reference modality are selectively supplemented to the target modality through Cross Mamba.
In the SFF module of the frequency domain, we perform element-wise summation for the phase spectrum and selective integration for the amplitude spectrum, as the phase spectrum of both modalities primarily contains consistent structural information while the amplitude spectrum from different modalities holds incompatible style information.
Finally, we adopt the ASFF module to enhance the fused features from both domains, where less informative channels from one domain are supplemented by incorporating the corresponding channel features from the other domain. The ASFF module enables the integration of relevant information and the suppression of redundant features.
Our contributions can be summarised as follows:
{\begin{itemize}
  \item We propose MMR-Mamba, an efficient framework for multi-modal MRI reconstruction. To our knowledge, this is the first exploration of Mamba to integrate complementary information across multi-modal MR images.

  \item We design the TCM module in the spatial domain for complementary feature fusion and the SSF module in the frequency domain for global structure information fusion.
  \item We introduce the ASFF module for spatial-frequency information fusion, enhancing task-relevant features while suppressing irrelevant features from these two domains.
  \item Extensive experiments on the BraTS and fastMRI datasets validate the effectiveness of our MMR-Mamba framework, demonstrating its superior performance compared to existing methods.
\end{itemize}

\section{Related Works}
\label{Rews}

\subsection{Spatial Domain MRI Reconstruction}
Benefiting from the complementary information from multi-modal images, many methods have been proposed to reconstruct high-quality images from low-quality images in the spatial domain, guided by the reference modality images\cite{zhou2023dsformer,xiang2018deep,feng2022multi,wang2023md}.
Previously, Xiang \etal \cite{xiang2018deep} proposed Dense-Unet to reconstruct T2WIs from T1WIs and under-sampled T2WIs, with the concatenation of under-sampled T2WIs and T1WIs serving as the input of the network. Xuan et al. \cite{xuan2022multimodal} further enhanced performance by introducing a spatial alignment network to compensate for spatial misalignment between multiple modality images.

Recently, due to the Transformer's capability in modeling long-range dependencies, several works have explored Transformer-based approaches for multi-modal MRI reconstruction.
MTrans \cite{feng2022multi} employed a cross-attention module for multi-scale feature fusion of the target modality and the reference modality. MD-GraphFormer \cite{wang2023md} incorporated the physical constraint of MRI into the network architecture and modeled the multiple contrasts as graph nodes for the joint reconstruction of multi-modal MR images over the graph.
MCCA \cite{li2023multi} introduced the hybrid convolutional transformer operation to enrich global and local spatial context representation. 
DCAMSR \cite{huang2023accurate} proposed a dual cross-attention mechanism, where the features of the reference image and the upsampled input image are extracted and promoted with both spatial and channel attention at multiple resolutions.
Despite the promising performance, the CNN and Transformer-based approaches either fall short of fully capturing long-range dependencies or suffer from quadratic complexity. In contrast, our model achieves comprehensive exploration of long-range dependencies without imposing a heavy computational burden.

\subsection{Dual-Domain MRI Reconstruction}
Given that aliasing artifacts in the images are structural and non-local, relying solely on image domain restoration is insufficient to eliminate them and achieve faithful reconstruction results. Consequently, several studies have explored the utilization of both k-space information and spatial domain information for MRI reconstruction  \cite{yang2017dagan,mirza2023learning,zhou2020dudornet,wang2023dsmenet}. 
For example, Yang \etal \cite{yang2017dagan} incorporated frequency domain information as additional constraints, formulated as an extra loss term. Similarly, the Fourier-constrained diffusion bridge (FDB) was introduced for diffusion model-based MRI reconstruction \cite{mirza2023learning}.
Subsequently, DuDoRNet \cite{zhou2020dudornet} proposed a novel paradigm that simultaneously recovers k-space and images to accelerate MR imaging through Residual Dense Network (DRDNet). 
Building on DuDoRNet, DuDoCAF \cite{lyu2022dudocaf} adopted a recurrent transformer structure to fuse features from two modalities for cross-modality reconstruction.
Liu \etal \cite{liu2024image} proposed Faster Fourier Convolution (FasterFC) for 3D MRI reconstruction,  directly restoring frequency domain information.
Furthermore, FMTNet \cite{yi2023frequency} separately repaired frequency information through a high-frequency learning branch and a low-frequency learning branch, and the two branches are concatenated for final results.
McSTRA \cite{ekanayake2024mcstra} also designed low-pass and high-pass reconstruction branches for frequency information restoration, 
and the whole reconstruction iterates between intermediate de-aliasing and data consistency via cascaded Swin-Transformers.

However, current dual-domain methods typically use similar networks for both k-space and image domains, overlooking the specific characteristics inherent to each domain. Moreover, these methods often rely on straightforward parallel or sequential combinations of two domains, failing to fully exploit cross-domain complementary information.

\subsection{State Space Models}
State Space Models (SSMs) offer a powerful framework for efficiently modeling long sequences and have garnered significant attention \cite{hasani2022liquid,gu2022train}. 
Previously, Structured State Space Sequence Models (S4) \cite{gu2021efficiently} was designed to capture long-range dependencies within sequences by introducing Higher-Order Polynomial Project Operator \cite{gu2020hippo}. 
S4 has demonstrated exceptional performance across various benchmarks while reducing computational complexity to $O(Nlog(N))$, significantly superior to the quadratic computational complexity $O(N^2)$ of Transformers. 
More recently, S6, \ie,~Mamba \cite{gu2023mamba}, was introduced. It employs a selection mechanism for choosing relevant information based on input and features an efficient hardware-aware algorithm grounded in selective scanning.

Mamba has significantly advanced natural language tasks, surpassing traditional Transformers.
Subsequently, Mamba has been extended to various vision tasks. For instance, 
Visual Mamba (Vim) \cite{zhu2024vision} proposed a pure Mamba-based backbone utilizing bidirectional Mamba blocks, modeling images in a sequence manner through positional embedding, akin to ViT \cite{dosovitskiy2020image}.
Visual State Space Model (Vmamba) \cite{liu2024vmamba} introduces a cross-scanning mechanism 
for spatial traversal, converting non-causal visual images into ordered patch sequences.
In the medical imaging domain, Mamba has been explored for tasks such as medical image segmentation \cite{xing2024segmamba,ma2024u}, registration \cite{guo2024mambamorph}, and classification \cite{yue2024medmamba,yang2024cmvim}.
For MRI reconstruction, Huang \etal \cite{huang2024mambamir} first proposed MambaMIR with an Arbitrary-Mask mechanism for medical image reconstruction.  
However, the complementary information from multi-modal images is neglected.
In this work, we explore Mamba for the integration of features from different modalities, harnessing its capability in efficiently modeling long-range dependencies.

%
% \vspace{-2pt}
\section{Methodology}
\label{Method}

\begin{figure*}[!ht]
\centering
\setlength{\abovecaptionskip}{4pt}
\setlength{\belowcaptionskip}{0pt}
\includegraphics [width=0.99\textwidth]{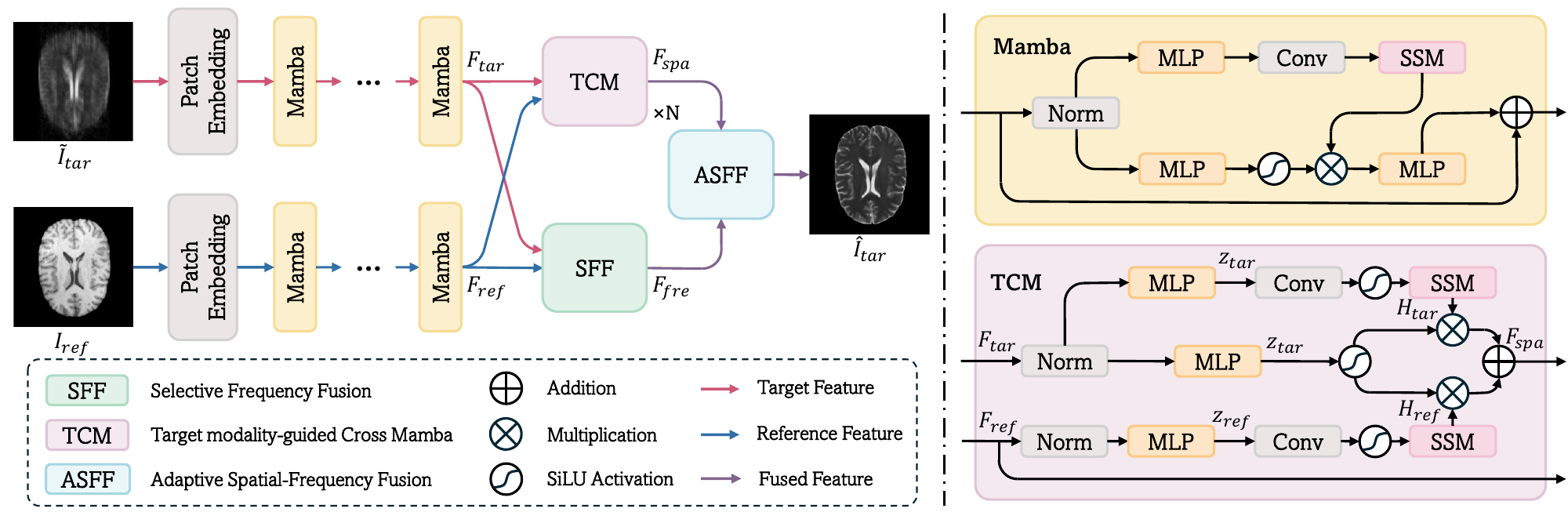}
\caption{\textbf{Overview of the proposed MMR-Mamba framework} (left). It contains Mamba blocks for feature extraction, TCM for spatial domain fusion, SFF for frequency domain fusion, and ASFF for spatial-frequency information integration. \textbf{Structure of Mamba block and TCM} (right).}
\label{FigOverview}
\vspace{-4pt}
\end{figure*}

\subsection{Preliminaries}
\vspace{3pt}\noindent\textbf{State Space Models.} 
SSMs are typically defined as linear, time-invariant systems that map an input sequence $x(t)\in\mathbb{R}^{L}$ to an output sequence $y(t)\in\mathbb{R}^{L}$ through a hidden state $h(t)\in\mathbb{R}^{N}$. These systems can be mathematically expressed as the following ordinary differential equation (ODE):
\begin{equation}
\begin{aligned}
h^{\prime}(t) = A h(t) + B x(t), \\
y(t) = C h(t) + D x(t),
\label{ssm1}
\end{aligned}
\end{equation}
where $A\in\mathbb{R}^{N\times N}$ denotes the state matrix,  $B\in\mathbb{R}^{N\times 1}$ and $C\in\mathbb{R}^{1\times N}$ represent the projection parameters, and $D\in\mathbb{R}^{1}$ is a skip connection.

To incorporate SSMs into deep learning algorithms, discretization is indispensable. The system in Eq \ref{ssm1} is discretized through the zero-order hold (ZOH). After discretization, the system can be written as

\begin{equation}
\begin{aligned}
& h_k = \overline{A} h_{k-1} + \overline{B} x_k, \\
& y_k = \overline{C} h_k + \overline{D} x_k,
\label{ssm2}
\end{aligned}
\end{equation}
where $\overline{A}=exp(\Delta A)$, $\overline{B}=(\Delta A)^{-1}(exp(\Delta A)-I)\cdot \Delta B$, $\overline{C}=C$, $\overline{D}=D$ are discretized parameters, and $\Delta$ is the discretization step size, which can be seen as the resolution of the continuous input $x(t)$.

Furthermore, the Eq. \ref{ssm2} can also be reformulated and computed as the following convolution:

\begin{equation}
\begin{aligned}
& \overline{K}=(C\overline{B}, C\overline{AB},..., C\overline{A}^{L-1}\overline{B}), \\
& y=x*\overline{K},
\label{ssm3}
\end{aligned}
\end{equation}
where $L$ denotes the length of the input sequence $x$ and $K \in \mathbb{R}^{L}$ is the SSM convolution kernel.

\vspace{3pt}\noindent\textbf{Fourier Transform.} Here, we briefly review the meaning of the Fourier transformation of images for a better understanding of our work.
The Fourier transform (FT) serves as a crucial technique in analyzing the frequency characteristics of an image. Transforming images from the spatial domain to the frequency domain through FT allows us to examine the images from a global perspective.
Given an image $x$, the FT can be expressed as follows:
\begin{align}
    \mathcal{F}(x)(u, v) = \sum_{h=0}^{H-1} \sum_{w=0}^{W-1} x(h,w) e^{-j2\pi(\frac{h}{H}u + \frac{w}{W}v)},
\end{align}
where $u$ and $v$ are coordinates in the Fourier space.
The frequency domain feature $\mathcal{F}(x)$ is represented as $\mathcal{F}(x) = \mathcal{R}(x) + j\mathcal{I}(x)$, with 
$\mathcal{R}(x)$ and $\mathcal{I}(x)$ denote the real and imaginary part respectively. 
Then, the amplitude spectrum $\mathcal{A}(x)(u,v)$ and phase spectrum $\mathcal{P}(x)(u,v)$ are defined as:
\begin{equation}
\begin{aligned}
    & \mathcal{A}(x)(u,v) = {\left[\mathcal{R}^2(x)(u,v) + \mathcal{I}^2(x)(u,v) \right] } ^{\frac{1}{2}}, \\
    & \mathcal{P}(x)(u,v) = arctan \left[ \frac{\mathcal{I}(x)(u,v)}{\mathcal{R}(x)(u,v)}\right].
\end{aligned}
\end{equation}

According to the Fourier theory, the amplitude spectrum $\mathcal{A}$ encodes the style information of an image, whereas the phase component spectrum $\mathcal{P}$ characterizes the structure information \cite{xu2021fourier,yang2020fda}.
Furthermore, as highlighted in \cite{yu2022frequency,zhou2022spatial}, the phase spectrum maintains global structure well, and the degradation of images is mainly manifested in the amplitude spectrum.

\subsection{Overall Architecture}
Let ${I}_{tar}$ and ${I}_{ref}\in\mathbb{R}^{H\times W}$ be a pair of MR images from the target modality and reference modality. Our objective is to reconstruct the high-quality target modality image $I_{tar}$ through a network that takes the under-sampled target modality image $\widetilde{I}_{tar}$ and the fully-sampled reference modality image ${I}_{ref}$ as inputs. Here,  $\widetilde{I}_{tar} = \mathcal{F}^{-1}(\mathcal{M} \odot k_{tar})$, where $\mathcal{F}^{-1}$ represents the inverse Fourier transform (IFT), $\mathcal{M}$ denotes the binary under-sample mask, and $k_{tar}$ represents the fully-sampled k-space data of the target modality. %with $\mathcal{F}^{-1}$ represent the inverse Fourier transform (IFT). 
The multi-modal reconstruction task can be formulated as the following optimization problem:
\begin{align}
  \hat{I}_{tar} = \argmin_{f(\theta)}\|I_{tar} - f(\mathcal{F}^{-1}(\mathcal{M} \odot k_{tar}),I_{ref};\theta)\|_1.
\label{Eq1}
\end{align}

To fully harness and fuse the information from different modalities, we design the MMR-Mamba network, depicted in Fig. \ref{FigOverview}. Initially, we adopt CNN blocks $\psi$ for low-level feature extraction and Mamba blocks $\phi$ for high-level feature extraction, producing modality-specific features $F_{tar}$ and $F_{ref}$ for the target modality and reference modality, respectively.
\begin{equation}
    \begin{aligned}
        F_{tar} = \phi_{tar} (\psi_{tar} (\widetilde{I}_{tar})), ~F_{ref} = \phi_{ref} (\psi_{ref} ({I}_{ref})).
    \end{aligned}
\end{equation}
Subsequently, we design the \textit{Target modality-guided Cross Mamba} (TCM) module in the spatial domain and the \textit{Selective Frequency Fusion} (SFF) module in the frequency domain, generating domain-specific fused features $F_{spa}$ and $F_{fre}$:
\begin{equation}
    \begin{aligned}
        F_{spa} = TCM(F_{tar},F_{ref}), ~F_{fre} = SFF(F_{tar},F_{ref}). 
    \end{aligned}
\end{equation}
Lastly, the multi-modal features in these two domains are further integrated through an \textit{Adaptive Spatial-Frequency Fusion} (ASFF) module to generate the final fused feature, which is then fed into a CNN Decoder to reconstruct the target modality image:
\begin{align}
  \hat{I}_{tar} = Decoder(ASFF(F_{spa}, F_{fre})).
\end{align}
The model is trained by minimizing the $L_1$ loss between the reconstructed image and the ground-truth image:
\begin{align}
  \mathcal{L}=\frac{1}{N} \sum_{n=1}^N \|\hat{I}_{tar} - I_{tar}\|_1 ,
\label{Eq_loss}
\end{align}
where $N$ is the number of training samples.

\subsection{Target-guided Cross Mamba}

As correlated features distribute across different regions and modalities, capturing and fusing these correlated features is crucial for the reconstruction of high-quality target modality images.
% For the extracted target modality feature $F_{tar}$ and reference modality feature $F_{ref}$, complementary information is distributed across different regions, making efficient integration crucial for reconstructing high-quality images from under-sampled data.
%
Existing fusion methods based on CNNs and Transformers face challenges due to their restricted capability in handling long-range dependencies or their high computational requirements.
To address this, we leverage state space models for target and reference feature fusion, benefiting from their ability to efficiently model long-range dependencies. 
Inspired by the cross-modal Mamba in \cite{he2024pan}, we design the \textit{Target-guided Cross Mamba} (TCM) module to enrich the target modality features with complementary information from reference features.

Specifically, we derive $z_{tar}$ and $z_{ref}$ from the target modality feature $F_{tar}$ and reference modality feature $F_{ref}$ using a normalization layer and a linear layer:
\begin{equation}
    \begin{aligned}
        &z_{tar} = Linear(Norm(F_{tar})), \\
        &z_{ref} = Linear(Norm(F_{ref})).
    \end{aligned}
\end{equation}
Then $z_{tar}$ and $z_{ref}$  are projected into the hidden state space through one-dimensional convolution with the SiLU activation function and SSM without gating as
\begin{equation}
    \begin{aligned}
        &H_{tar} = SSM(SiLU (Conv1d(z_{tar}))), \\
        &H_{ref} = SSM(SiLU (Conv1d(z_{ref}))).
    \end{aligned}
\end{equation}
After that, to maximally restore the target modality information and selectively integrate the reference information, we adopt $z_{tar}$ as the gating to modulate hidden state features  $H_{tar}$ and $H_{ref}$, and implement the hidden state feature fusion as
\begin{equation}
    \begin{aligned}
        &F_{spa} = H_{tar} \otimes SiLU(z_{tar}) +  H_{ref} \otimes SiLU(z_{tar}), 
    \end{aligned}
\end{equation}
where $\otimes$ represents element-wise production.
In this way, we build the TCM module in a hidden state space based on the gating mechanism. The gating parameters are derived from the target modality, ensuring that the target modality predominates in the fusion process and selectively incorporates complementary information from the reference modality.

In our framework, we stack four TCM modules, incorporating residual connections to preserve the integrity of target features. The fused feature $F_{spa}$ is subsequently forwarded to the spatial-frequency fusion module to facilitate the reconstruction of target images.

% \begin{equation}
%     \begin{aligned}
%         \var{F_{fus}} = Linear(F_{fus})+F_{fre}
%     \end{aligned}
% \end{equation}

\subsection{Selective Frequency Fusion}

To further fuse the complementary information from the extracted features in a broader view, we resort to the frequency domain (\ie, ~Fourier domain), where each frequency component corresponds to all the pixels in the spatial domain, inherently capturing global properties.
Additionally, as highlighted in \cite{yu2022frequency,zhou2022spatial}, the phase spectrum in the Fourier domain preserves global structure well, while the image degradation mainly manifests in the amplitude spectrum.
To comprehensively integrate the complementary information and to restore the degraded features in the amplitude spectrum, we propose the \textit{Selective Frequency Fusion} (SFF) module within the frequency domain, illustrated in Fig. \ref{FigFreFusion}.

Given the extracted features $F_{tar}$ and $F_{ref}$ from previous modules, we first transform them to frequency domain through Fourier transform and obtain their amplitude spectrum and phase spectrum:
\begin{equation}
\begin{aligned}
    \mathcal{A}_{tar}, \mathcal{P}_{tar} &= \mathcal{F}(F_{tar}), \\
    \mathcal{A}_{ref}, \mathcal{P}_{ref} &= \mathcal{F}(F_{ref}).
\end{aligned}
\end{equation}
For the phase spectrum fusion, we perform element-wise addition on  $\mathcal{P}_{ref}$ and $\mathcal{P}_{tar}$, as both contain crucial and consistent structural information \cite{skarbnik2009importance,liu2021feddg}. 
%
%To simplify the notation, we omit any $(x)(u,v)$ on $\mathcal{A}(x)(u,v)$ and $\mathcal{P}(x)(u,v)$ in following description.
%
Regarding the amplitude spectrum, it encapsulates style information, which varies significantly across different modalities. Moreover, the amplitude of the under-sampled low-quality images contains interference information that can negatively impact the final reconstruction.
Direct concatenation of the amplitude spectrum from these modalities would introduce incompatible and interference information, compromising the quality of the reconstructed image \cite{skarbnik2009importance,liu2021feddg}.

To mitigate this issue, we adopt a selective strategy to fuse the amplitude spectrum.
Specifically, we design a \textit{Selective Amplitude Harmonization Module} (SAHM), wherein the amplitude spectrum across modalities is dynamically adjusted based on global statistics.
First, we conduct element-wise addition of $\mathcal{A}_{ref}$ and $\mathcal{A}_{tar}$ to produce an intermediate feature $\mathcal{A}$. 
Next, we adopt global average pooling to generate channel-wise statistics $s \in\mathbb{R}^C $, embedding global information. Specifically, the $c-th$ element of $s$ is calculated by shrinking $\mathcal{A}$ through spatial dimensions $H \times W$.
% \begin{align}
%     s_c=f_{gp}(\mathcal{A}_c)=\frac{1}{H \times W}\sum_{i=1}^H\sum_{j=1}^W \mathcal{A}_c(i,j).
% \end{align}
For computational efficiency, a fully connected (FC) layer is utilized to create a compact feature $z\in \mathbb{R}^{L\times1}$, which is further used to guide the adaptive selection. The operation can be expressed as follows:
\begin{align}
    z=f_{fc}(s)=\sigma(\mathcal{B}(\mathbf{W}s),
\end{align}
where $\sigma$ is the ReLU activation function, $\mathcal{B}$ represents batch normalization, and $\mathbf{W}\in \mathcal{R}^{L\times C}$ with $L$ means the number of channel after dimension reduction.

To dynamically select different scales of amplitude spectrum from the two modalities, the compact feature descriptor  $z\in \mathbb{R}^{L\times1}$ is used to compute a soft attention score across channels. To achieve this, softmax is performed on the channel-wise digits:
\begin{equation}
\begin{aligned}
    a_c=\frac{e^{A_cz}}{e^{A_cz} + e^{B_cz}}, ~
    b_c=\frac{e^{B_cz}}{e^{A_cz} + e^{B_cz}},
\end{aligned}    
\end{equation}
where $A,B\in \mathcal{R}^{C\times L}$ and $a,~b$ represent the soft attention vector for 
$\mathcal{A}_{ref}$ and $\mathcal{A}_{tar}$. Then, the fused amplitude spectrum $\overline{\mathcal{A}}$ is obtained by weighted sum of $\mathcal{A}_{ref}$ and $\mathcal{A}_{tar}$:
\begin{align}
    \overline{\mathcal{A}_c} = a_c \cdot \mathcal{A}_{ref} + b_c \cdot \mathcal{A}_{tar}.
\end{align}

\begin{figure}[!t]
\centering
\setlength{\abovecaptionskip}{4pt}
\setlength{\belowcaptionskip}{0pt}
\includegraphics [width=0.5\textwidth]{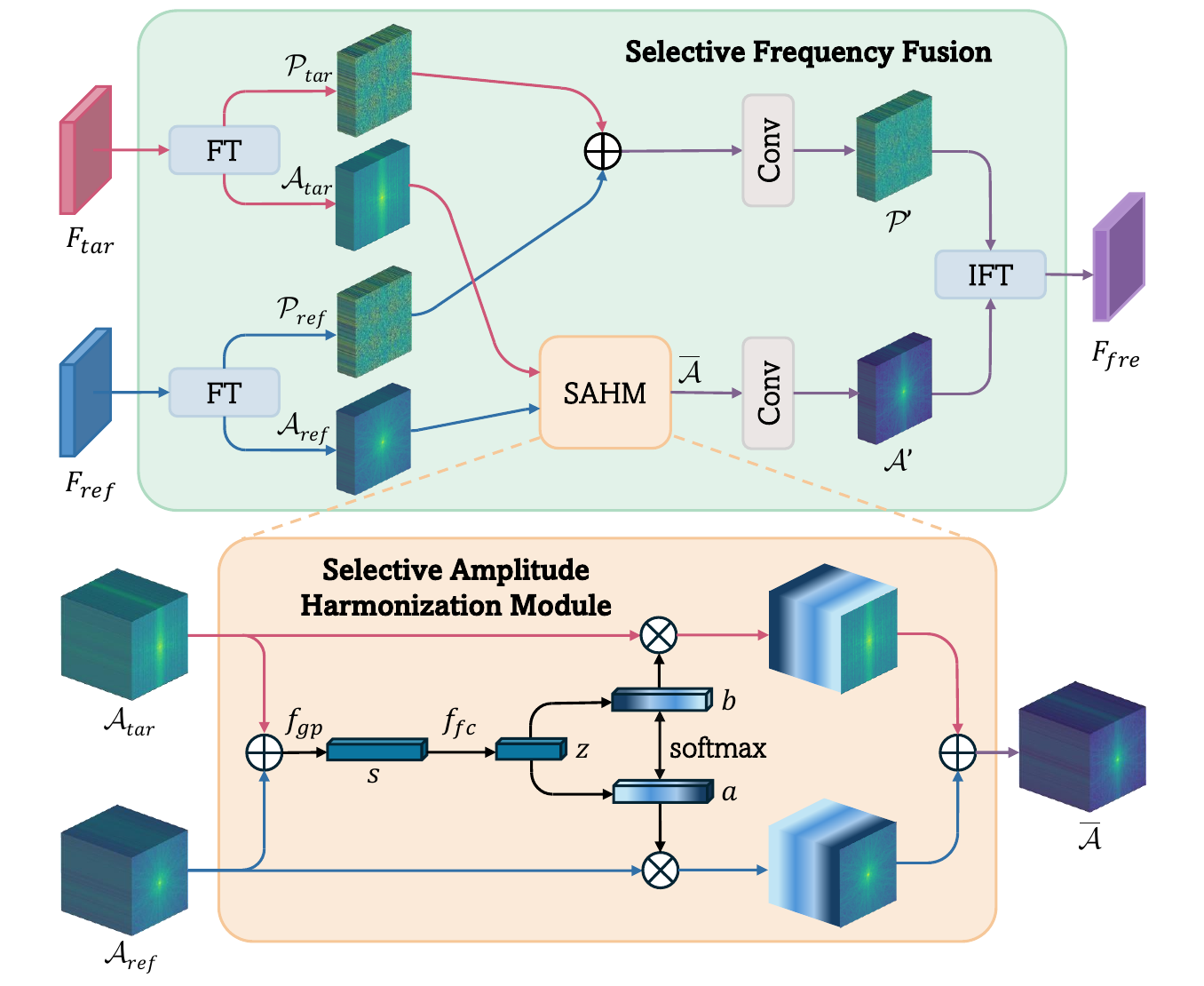}
\caption{Illustration of Selective Frequency Fusion (SFF) module.}
\label{FigFreFusion}
\vspace{-4pt}
\end{figure} 

To further enhance the fused amplitude spectrum and phase spectrum, we employ two groups of independent operations $ConvA(\cdot)$ and $CovnP(\cdot)$. Each group consists of convolution layers with a kernel size of $1\times1$ and a ReLU activation function. The final restored amplitude spectrum $\mathcal{A}^\prime$ and phase spectrum $\mathcal{P}^\prime$ is obtained by
\begin{equation}
\begin{aligned}
    &\mathcal{A}^\prime = ConvA(\overline{\mathcal{A}}),\\
    &\mathcal{P}^\prime = ConvP(\mathcal{P}_{ref} + \mathcal{P}_{tar}).
\end{aligned}
\end{equation}
    
Finally, the restored amplitude spectrum $\mathcal{A}^\prime$ and phase spectrum $\mathcal{P}^\prime$ are convert to spatial domain through inverse Fourier transform:
\begin{align}
    F_{fre} = \mathcal{F}^{-1}(\mathcal{A}^\prime,\mathcal{P}^\prime),
\end{align}
where $\mathcal{F}^{-1}$ donates inverse Fourier transform (IFT).
Through the FT and selective fusion in the Fourier domain, the SFF module efficiently achieves global feature fusion and recovers high-frequency signals for reconstructing structural details.

\begin{figure}[!t]
\centering
\setlength{\abovecaptionskip}{4pt}
\setlength{\belowcaptionskip}{0pt}
\includegraphics [width=0.5\textwidth]{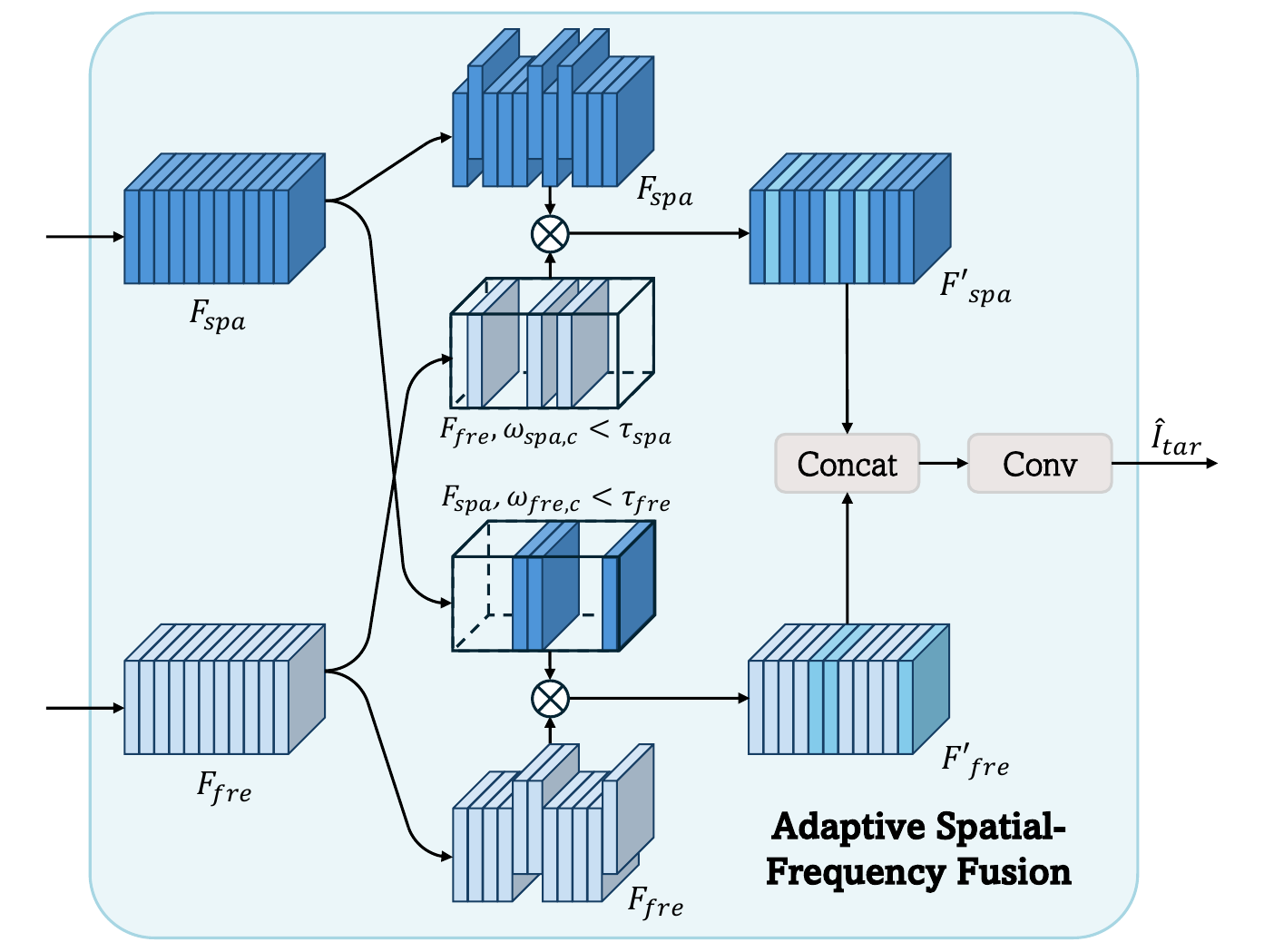}
\caption{Illustration of the Adaptive Spatial-Frequency Fusion module.}
\label{FigDualFusion}
\vspace{-4pt}
\end{figure}

\subsection{Adaptive Spatial-Frequency Fusion}
To further enhance the fused spatial domain feature $F_{spa}$ and fused frequency domain feature $F_{fre}$, and to facilitate the efficient integration of these features, we introduce the \textit{Adaptive Spatial-Frequency Fusion} (ASFF) module, as illustrated in Fig. \ref{FigDualFusion}. 
The ASFF module allows the two domains to mutually complement each other through
channel-wise adaptive integration based on the informativeness of each channel, which is measured by the Batch-Normalization (BN) layer.
%
% According to previous work~\cite{liu2017learning,ye2018rethinking}, the BN value serves as an indicator of channel importance; lower norms signify lesser relevance to the ultimate outcomes.
Previous studies~\cite{liu2017learning,ye2018rethinking} have shown that the  BN value indicates channel importance, with lower norms signifying lesser relevance to the final outcomes.
Consequently, we propose that channels with diminished significance in one domain can be reinforced by incorporating corresponding channel information from the other domain. 

Firstly, we measure the channel-wise informativeness of the fused spatial domain feature $F_{spa}$ and frequency feature $F_{fre}$ through the BN layer. 
Denote the $c-th$ channel of the features as $F_{spa,c}$ and $F_{fre,c}$, the BN is computed as follows:
\begin{equation}
\begin{aligned}
     F_{spa,c}^\prime &= \omega_{spa,c} \frac{F_{spa,c}-\mu_{spa,c}}{\sqrt{\sigma_{spa,c}^2 + \epsilon}} + \beta_{spa,c}, \\
     F_{fre,c}^\prime &= \omega_{fre,c} \frac{F_{fre,c}-\mu_{fre,c}}{\sqrt{\sigma_{fre,c}^2 + \epsilon}} + \beta_{fre,c},
\end{aligned}
\end{equation}
where $\omega_{spa,c}$ and $\omega_{fre,c}$ are trainable scaling factors; $\beta_{spa,c}$ and $\beta_{fre,c}$ are trainable
offsets, and $\epsilon$ is a small constant to avoid division by zero.

The factors $\omega_{spa,c}$ and $\omega_{fre,c}$ evaluate the correlation between the input and the output after normalization during training. 
If $\omega_{tar,c}$ or $\omega_{fre,c}$ approaches 0,  it indicates that the corresponding gradient loss of $F_{spa}$ or $F_{fre}$ will be close to 0. This implies that $ F_{spa,c}^\prime$ or $F_{fre,c}^\prime$ will have minimal influence on the final results.
Given this insight, we propose to enhance the channels with small scaling factors in one domain by incorporating the corresponding channels from the other domain. The incorporation is performed as follows:
\begin{equation}
F_{spa,c}^\prime = \left\{  
\begin{aligned}
            & F_{spa,c},  &if ~ \omega_{spa,c}~ \geq ~\tau_{spa};  \\  
            & F_{spa,c} \otimes F_{fre,c},    &if~ \omega_{spa,c}~ \textless ~\tau_{spa};
\end{aligned}
\right. 
\end{equation}
\begin{equation}
F_{fre,c}^\prime = \left\{  
\begin{aligned}
            & F_{fre,c},  &if ~ \omega_{fre,c}~ \geq ~\tau_{fre};  \\  
            & F_{fre,c} \otimes F_{spa,c},   &if~ \omega_{fre,c}~ \textless ~\tau_{fre};
\end{aligned}
\right.
\end{equation}
where $\otimes$ denotes element-wise multiplication and $\tau_{spa}$ and $\tau_{fre}$ are the thresholds, set according to the maximum and the minimum of the scaling factor as follows:
\begin{equation}
\begin{aligned}
    \tau_{spa} = \omega_{spa}^{min} + \alpha (\omega_{spa}^{max}-\omega_{spa}^{min}), \\
    \tau_{fre} = \omega_{fre}^{min} + \alpha (\omega_{fre}^{max}-\omega_{fre}^{min}), 
\end{aligned}
\end{equation}
where  $\alpha$ is a value set to 0.1 in our experiments.

Through the ASFF module, less informative features from one domain are enhanced by the corresponding features from the other, while redundant information is effectively eliminated. This process ensures that each channel contributes optimally to the final fused features. 
By leveraging the most relevant information from both spatial and frequency domains, the ASFF module produces feature representations that are most conducive to generating high-quality target images. 
% This adaptive integration strategy improves the overall efficiency and effectiveness of feature fusion, leading to enhanced reconstruction performance.
%
After the channel-wise incorporation, the features $F_{spa}^\prime$ and $F_{fre}^\prime$ are concatenated and fed to the CNN Decoder to recover the reconstructed target modality image $\hat{I}_{tar}$:
\begin{align}
    \hat{I}_{tar} = Decoder(Cat\left[ F_{spa}^\prime, F_{fre}^\prime \right]).
\end{align}

%%%%%%%%%%%%%%%%%%%%%%%%%%%%%%%%%%%%%%%%%%%%%%%%%
\section{Experiments and Results}
\label{Exp}

\subsection{Dataset Description}
In this study, we adopt two datasets with different anatomical structures and protocols for evaluation: the BraTS dataset \cite{menze2014multimodal} and the fastMRI knee dataset \cite{knoll2020fastmri}.
\textbf{The BraTS Dataset} contains both T1WIs and T2WIs scans of the brain. We extract 2D images uniformly from 100 3D MRI volumes in the BraTS dataset. The dataset is split subject-wise with a 3:1 ratio, yielding a total of 3,621 images for training and 1,088 images for testing. The 2D image size is $240 \times 240$. In our experiments, we use T1WI as the reference modality for the reconstruction of the T2WI modality.
\textbf{The fastMRI Dataset} is the largest public MRI  dataset with raw k-space data. Following \cite{xuan2020learning}, we select 227 and 45 pairs of single-coil PDWI and FS-PDWI knee volumes for training and testing, respectively, resulting in a total of 8,332 pairs of 2D images for training and 1,665 images for testing. The 2D image size is $320 \times 320$. In our experiments, we use PDWI as the reference modality for the reconstruction of the FS-PDWI modality.

\subsection{Experimental Setup}
\label{setup}
\vspace{0pt}\noindent\textbf{Comparison Methods.} 
To demonstrate the effectiveness of our proposed MMR-Mamba, we compare it against seven multi-modal reconstruction methods: MDUNet~\cite{xiang2018deep}, MINet~\cite{feng2021multi}, MCCA~\cite{li2023multi}, MTrans~\cite{feng2022multi}, DCAMSR~\cite{huang2023accurate}, SwinIR~\cite{liang2021swinir}, and Pan-Mamba~\cite{he2024pan}.
For a fair comparison, we run all the models for $100,000$ iterations with a batch size of 4.
Other settings of the compared methods followed the implementations provided in their original papers.

\vspace{0pt}\noindent\textbf{Performance Metrics.} 
For quantitative evaluation, we assess the image reconstruction results using 
peak signal-to-noise ratio (PSNR), structural similarity index (SSIM), and normalized mean squared error (NMSE). For qualitative evaluation, we visualize the reconstruction results with error maps.

\vspace{0pt}\noindent\textbf{Implementation Details.} The proposed framework is developed with PyTorch, and the training and testing processes are executed on an NVIDIA RTX A6000 GPU (48GB).
We employ an AdamW optimizer with an initial learning rate of $0.001$ and a weight decay of $1e-4$ for the training. 
We run $100,000$ iterations with a batch size of 4 for both datasets.
For both datasets, the undersampled images are obtained by applying a 1D Cartesian random under-sample mask with $4\times$ and $8\times$ acceleration.

\begin{table*}[t]
    \centering
    % \footnotesize
    \caption{
    \textbf{Quantitative results on the BraTS and fastMRI datasets with different acceleration factors.} We report mean±std for the PSNR, SSIM, and NMSE metrics, along with network parameters. The best results are highlighted in \colorbox{c1}{\textbf{red}}.}
    % \scriptsize
    \begin{tabular}{ll|ccc|ccc|c}
    \toprule
    % \rowcolor[HTML]{EFEFEF}
     \textbf{BraTS} & &\multicolumn{3}{c|}{$4 \times$} & \multicolumn{3}{c|}{$8 \times$}  &\multirow{2}{*}{Param (M)} \\
     % \midrule
    \cmidrule(lr){1-2} \cmidrule(lr){3-5} \cmidrule(lr){6-8} 
     Method & Year & PSNR$\uparrow$ & SSIM$\uparrow$ & NMSE~$(10^{-2})$$\downarrow$  & PSNR$\uparrow$ & SSIM$\uparrow$ & NMSE~$(10^{-2})$$\downarrow$ &  \\
     \midrule
     Zero-filling~\cite{bernstein2001effect} & JMRI'01 & 30.11±1.53 &0.767±0.038  & 5.659±1.816 & 26.58±1.49 & 0.673±0.036 & 12.469±3.337 & -\\
     MDUNet~\cite{xiang2018deep} & TBME'18 &37.94±1.66 &0.975±0.006 & 0.905±0.310 &35.19±1.64 &0.960±0.009 & 1.690±0.519 & 7.40\\
     MINet~\cite{feng2021multi} & MICCAI'21 & 38.26±1.74 &0.976±0.006 & 0.847±0.312 &35.23±1.72 &0.961±0.009 & 1.697±0.568 & 3.75\\
     SwinIR~\cite{liang2021swinir} & CVPR'21 & 37.87±1.73 &0.974±0.006 & 0.926±0.334 &34.95±1.72 &0.960±0.009 & 1.805±0.590 & 2.99\\
     MTrans~\cite{feng2022multi} & TMI'22 & 36.02±1.67 &0.962±0.007 & 1.429±0.496 &34.81±1.57 &0.957±0.009 & 1.868±0.574 & 746.02\\
     MCCA~\cite{li2023multi} & JBHI'23 & 38.03±1.68 &0.975±0.006 & 0.903±0.312 &35.37±1.66 &0.962±0.009 & 1.633±0.511 & 15.21\\
     DCAMSR~\cite{huang2023accurate} & MICCAI'23 & 38.60±1.75 & {0.978±0.006} & 1.193±0.402 &35.99±1.74 &0.965±0.009 & 1.417±0.464 & 10.83\\
     Pan-Mamba~\cite{he2024pan} & Arxiv'24 & {38.84±1.79} & {0.978±0.006} & {0.739±0.273} & {36.18±1.77} & {0.966±0.009} & {1.363±0.466} &2.12\\
     MMR-Mamba (Ours) & — & \cellcolor{c1}\textbf{40.98±1.88} & \cellcolor{c1}\textbf{0.985±0.005} & \cellcolor{c1}\textbf{0.454±0.190} & \cellcolor{c1}\textbf{37.75±1.85} & \cellcolor{c1}\textbf{0.974±0.008} & \cellcolor{c1}\textbf{0.955±0.358} & \cellcolor{c1}\textbf{0.42}\\

    \toprule
    \toprule
    \textbf{fastMRI} & &\multicolumn{3}{c|}{$4 \times$} & \multicolumn{3}{c|}{$8 \times$} &\multirow{2}{*}{Param (M)}  \\
    % \midrule
    \cmidrule(lr){1-2} \cmidrule(lr){3-5} \cmidrule(lr){6-8} 
     Method & Year & PSNR$\uparrow$ & SSIM$\uparrow$ & NMSE~$(10^{-2})$$\downarrow$ & PSNR$\uparrow$ & SSIM$\uparrow$ & NMSE~$(10^{-2})$$\downarrow$ &  \\
    \midrule
     Zero-filling~\cite{bernstein2001effect} & JMRI'01 & 27.68±1.75 &0.571±0.061 & 5.020±1.271 & 25.64±1.64 & 0.454±0.070 & 7.719±1.594 & -\\
     MDUNet~\cite{xiang2018deep} & TBME'18 &28.60±1.00 &0.600±0.050 & 4.000±1.021 &27.90±0.86 &0.544±0.050 & 4.600±1.942 & 7.40\\
     MINet~\cite{feng2021multi} & MICCAI'21 &29.47±1.88 &0.639±0.069 & 3.334±0.937 &28.17±1.74 &0.563±0.081 & 4.327±1.019 & 3.75\\
     SwinIR~\cite{liang2021swinir} & CVPR'21 &29.42±1.87 &0.636±0.069 & 3.382±0.949 &28.09±1.74 &0.560±0.081 & 4.407±1.032 &2.99\\
     MTrans~\cite{feng2022multi} & TMI'22 &29.00±1.79 &0.619±0.068 & 3.699±0.929 &27.31±1.68 &0.526±0.081 & 5.254±1.049 & 746.02\\
     MCCA~\cite{li2023multi} & JBHI'23 &29.46±1.87 &0.637±0.069 & 3.346±0.937 &28.23±1.75 &0.562±0.081 & 4.275±1.019 & 15.21\\
     DCAMSR~\cite{huang2023accurate} & MICCAI'23 &29.45±1.87 &0.637±0.068 & 3.349±0.935 & {28.42±1.79} &0.569±0.081 & {4.098±1.027} & 10.83\\
     Pan-Mamba~\cite{he2024pan} & Arxiv'24 & {29.59±1.87} & {0.645±0.068} & {3.248±0.928} & 28.36±1.76 & {0.570±0.081} & 4.159±1.034 & 2.12\\
     MMR-Mamba (Ours)  & — & \cellcolor{c1}\textbf{29.66±1.88} & \cellcolor{c1}\textbf{0.647±0.068} & \cellcolor{c1}\textbf{3.201±0.927} & \cellcolor{c1}\textbf{28.44±1.76} & \cellcolor{c1}\textbf{0.572±0.081} & \cellcolor{c1}\textbf{4.084±1.036} & \cellcolor{c1}\textbf{0.42}\\
    \bottomrule
    \end{tabular}
\label{tab:main_result}
\end{table*}

\subsection{Experimental Results}
\label{ExpRes}
In this section, we evaluate the proposed method against state-of-the-art techniques under various experimental settings.
% under various experimental settings: acceleration 4$\times$ and 8$\times$ on both datasets.

\vspace{0pt}\noindent\textbf{Quantitative Results.} 
In Table \ref{tab:main_result}, we report the PSNR, SSIM, and NMSE results on both datasets for 4$\times$ and 8$\times$ acceleration.
Firstly, focusing on the BraTS dataset in the upper part of the table, our method achieves the best reconstruction results, with a PSNR of 40.98 dB and SSIM of 0.985 under 4$\times$ acceleration, and a PSNR of 37.75 dB and SSIM of 0.974 under 8$\times$ acceleration. These results highlight our model's efficacy in fusing information from multi-modal images for reconstructing target images.
Additionally, our method outperforms the second-best method, Pan-Mamba, by 2.14 dB in PSNR under 4$\times$ acceleration and by 1.57 dB in PSNR under 8$\times$ acceleration. The fact that Pan-Mamba achieves the second-best results further highlights the effectiveness of the Mamba in MRI reconstruction. 
Similarly, as shown in the lower part of the table, our method achieves the best reconstruction results on the fastMRI knee dataset, with a PSNR of 29.66 dB and SSIM of 0.647 under 4$\times$ acceleration and a PSNR of 28.44 dB and SSIM of 0.572 under 8$\times$ acceleration. This further demonstrates the effectiveness and robustness of our model in reconstructing high-quality images.

\begin{table*}[t]
    \centering
    % \footnotesize
    \caption{
    \textbf{Ablation Study of the proposed modules on BraTS dataset.} We report mean±std for the PSNR, SSIM, and NMSE metrics. 
    }
    % \scriptsize
    \begin{tabular}{c|ccc|ccc|ccc}
    \toprule
    % \rowcolor[HTML]{EFEFEF}  \multirow{2}{*}{IXI}
     % & \textbf{BraTS}  &\multicolumn{3}{c}{$4 \times$} \\
     % \midrule
      \multirow{2}{*}{Model} & \multirow{2}{*}{TCM} & \multirow{2}{*}{SFF} & \multirow{2}{*}{ASFF} & \multicolumn{3}{c|}{4$\times$} & \multicolumn{3}{c}{8$\times$}\\
     \cmidrule(lr){5-7}\cmidrule(lr){8-10}
     &&&&PSNR$\uparrow$ & SSIM$\uparrow$ & NMSE~$(10^{-2})$$\downarrow$  & PSNR$\uparrow$ & SSIM$\uparrow$ & NMSE~$(10^{-2})$$\downarrow$ \\
     \midrule
      a & \XSolidBrush & \XSolidBrush & \XSolidBrush & 38.45+1.75 &0.977±0.006  & 0.809±0.294 &35.72±1.72 &0.964±0.009 & 1.512±0.499 \\
      b & \Checkmark & \XSolidBrush & \XSolidBrush & 39.05±1.79 &0.979±0.006 & 0.706±0.266  &36.17±1.80 &0.967±0.009 & 1.368±0.476 \\
      c & \XSolidBrush & \Checkmark & \XSolidBrush & 40.49±1.84 &0.984±0.005 & 0.509±0.205  &37.22±1.79 &0.971±0.008 & 1.078±0.392 \\
     % \midrule
      d & \Checkmark & \Checkmark & \XSolidBrush & 40.66+1.86 &0.985±0.005  & 0.489±0.201 &37.50±1.84 &0.973±0.008 & 1.012±0.374\\
     e & \Checkmark & \Checkmark & \Checkmark & \textbf{40.98±1.88} & \textbf{0.985±0.005} & \textbf{0.454±0.190} & \textbf{37.75±1.85} & \textbf{0.974±0.008} & \textbf{0.955±0.358} \\
    \bottomrule
    \end{tabular}
\label{tab:ablation_moule}
\end{table*}

\vspace{0pt}\noindent\textbf{Qualitative Results.}
To better evaluate the reconstruction quality, we visualize the outputs from different methods for the BraTS and fastMRI datasets under 4$\times$ and 8$\times$ acceleration in Fig. \ref{FigQualitative}, along with their corresponding error maps. These maps illustrate the discrepancies between the reconstructed and the ground truth images, with blue indicating minimal error and red indicating higher error.
The error maps reveal that zero-filled reconstructions exhibit pronounced artifacts,  with error levels escalating with increasing acceleration factors from 4$\times$ to 8$\times$. 
Notably, our method consistently shows the lowest reconstruction error across both datasets and acceleration factors. This observation underscores that our method ensures superior preservation of essential anatomical details and consistently produces high-quality reconstructed images.

\begin{figure*}[!ht]
\centering
\setlength{\abovecaptionskip}{4pt}
\setlength{\belowcaptionskip}{0pt}
\includegraphics [width=0.99\textwidth]{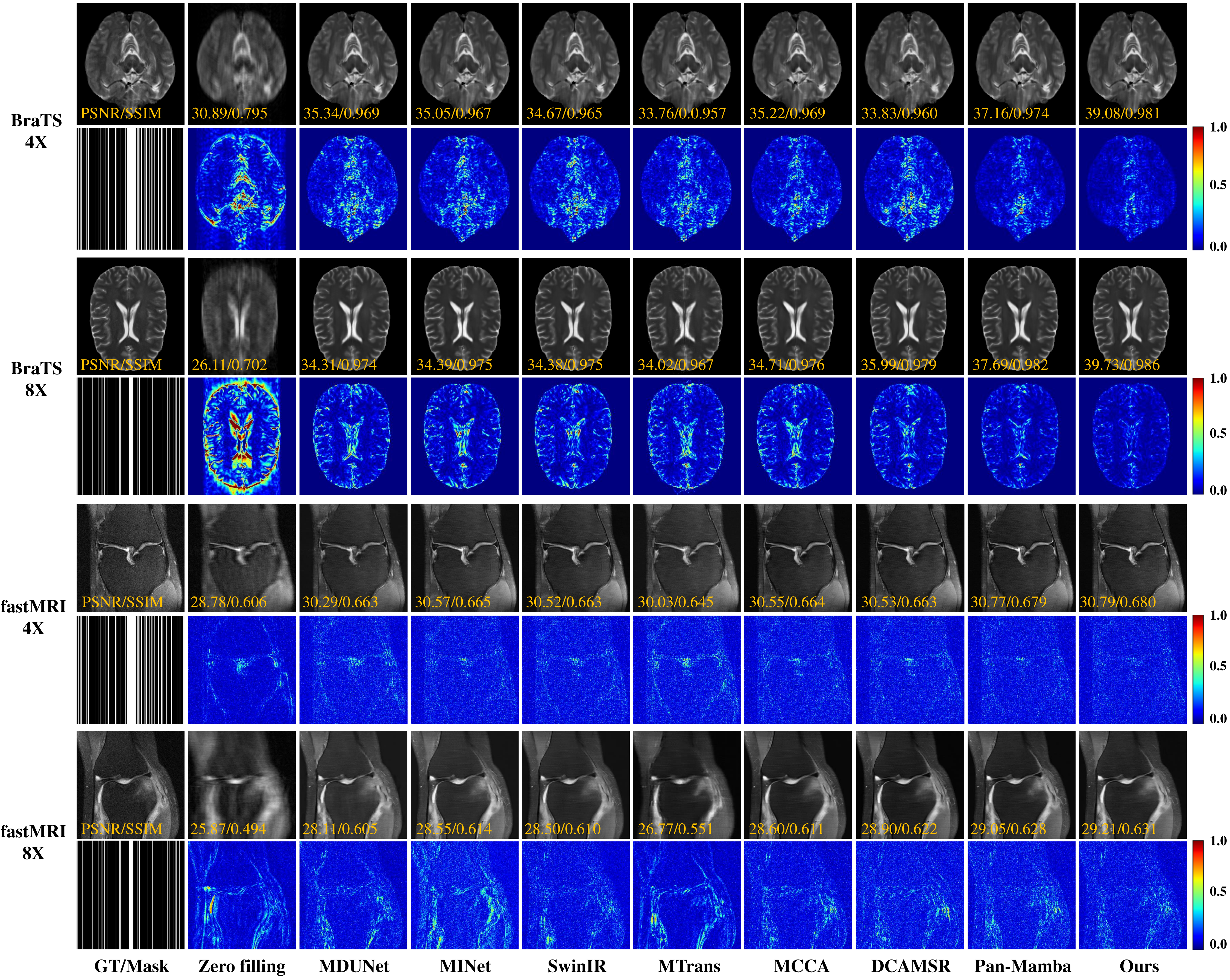}
\caption{Qualitative evaluation of reconstruction results from different methods on BraTS dataset and fastMRI knee dataset under 4$\times$ and 8$\times$ acceleration. For every group, the first row shows the reconstructed images and the second row displays the error map between the results and the ground truth. More color in the error map indicates worse reconstruction results.}
\label{FigQualitative}
\vspace{-4pt}
\end{figure*}

\subsection{Ablation Studies}
In this section, we first conduct an ablation study to assess the efficacy of the proposed modules. Then we analyze different fusion strategies for both spatial and frequency domains.

\subsubsection{Effectiveness of the Proposed Modules}
To verify the validity of the proposed modules, we conducted ablation experiments on the BraTS dataset under 4$\times$ and 8$\times$ acceleration. The different configurations tested are denoted as follows: (a) a baseline model with all proposed modules removed; (b) the baseline model enhanced with the TCM; (c) the baseline model enhanced with the SFF; (d) the model augmented with both the TCM and SFF; and (e) the complete model incorporating all proposed modules.
The results of these experiments are reported in Table \ref{tab:ablation_moule}. As shown, incorporating the TCM improves the PSNR from 38.45 dB to 39.05 dB under 4$\times$ acceleration, 
while incorporating the SFF improves the PSNR to 40.49 dB under 4$\times$ acceleration.                             
Combining both TCM and SFF modules further boosts PSNR to 40.66 dB. Finally, the inclusion of the ASFF led to additional improvements. Similar trends were observed under 8$\times$ acceleration.
Visual results on the BraTS dataset, including error maps, are shown in Fig. \ref{FigAblation}. As shown in the figure, the TCM module alone reduces the overall error, yet some structural details remain unrecovered, as indicated in the yellow box. Conversely, the SFF module effectively restores structural details but exhibits larger errors in the lateral ventricle areas, highlighted in the red box. When both the TCM and SFF modules are employed, most regions and structures are adequately restored. Ultimately, integrating all proposed modules yields the best results. 
The quantitative and visualization results from our ablation experiments verify the effectiveness of each proposed module in enhancing the performance of multi-modal MRI reconstruction.

\begin{figure*}[!t]
\centering
\setlength{\abovecaptionskip}{4pt}
\setlength{\belowcaptionskip}{0pt}
\includegraphics [width=0.85\textwidth]{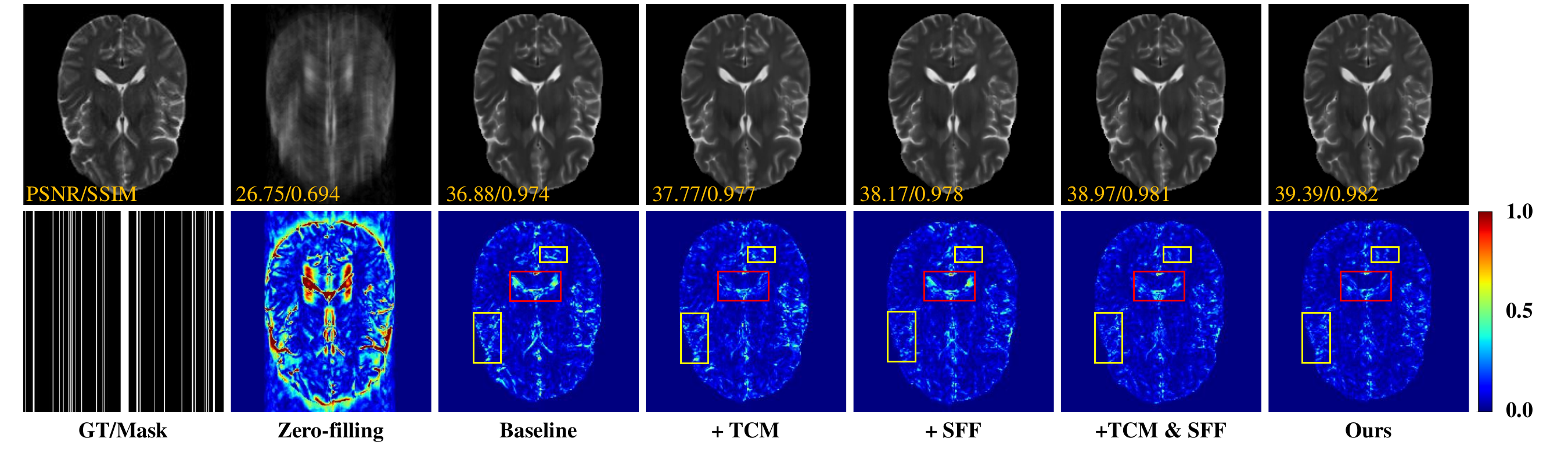}
\caption{Visualization of the results from ablation study of our proposed modules on BraTS dataset under 8$\times$ acceleration. }
\label{FigAblation}
\vspace{-4pt}
\end{figure*}

\subsubsection{Analysis on Spatial Domain Fusion} 
To evaluate the effectiveness of our proposed spatial domain fusion module, we design three ablation experiments: (1) fusion through element-wise addition, (2) fusion through cross-attention, and (3) fusion through our proposed TCM module, denoted as `Sum', `CA' and `TCM' in Table \ref{tab:ablation_spa}. 
The results indicate that element-wise addition of features from two modalities leads to the worst results. In contrast, our TCM module achieves superior performance, outperforming `Sum' and `CA' under both 4$\times$  and 8$\times$ acceleration, notably with minimal parameters.

\begin{table}[t]
    \centering
    % \footnotesize
    \caption{Ablation Study on spatial domain fusion on the BraTS dataset under $4\times$ and $8\times$ acceleration. 
    }
    % \scriptsize
    \begin{adjustbox}{width=0.5\textwidth}
    \begin{tabular}{c|c|ccc|c}
    \toprule
    % \rowcolor[HTML]{EFEFEF}  \multirow{2}{*}{IXI}
     % & \textbf{BraTS}  &\multicolumn{3}{c}{$4 \times$} \\
     % \midrule
     & Method & PSNR$\uparrow$ & SSIM$\uparrow$ & NMSE~$(10^{-2})$$\downarrow$ & Param(M) \\
     \midrule
     \multirow{2}{*}{$4\times$} 
     & Sum & 38.56+1.74 &0.977±0.006  & 0.789±0.285 & \textbf{0.28} \\
     & CA & 38.76+1.76 &0.978±0.006  & 0.754±0.278 & 1.48 \\
     & TCM & \textbf{39.05±1.79} & \textbf{0.979±0.006} & \textbf{0.706±0.266} & 0.39 \\
     \midrule
     \multirow{2}{*}{$8\times$} 
     & Sum & 35.82+1.74 &0.965±0.009  & 1.481±0.492  & \textbf{0.28}\\
     & CA & 35.92+1.74 &0.965±0.009  & 1.447±0.492 & 1.48\\
     & TCM  & \textbf{36.17±1.80} & \textbf{0.967±0.009} & \textbf{1.368±0.476} & 0.39\\
     % \midrule
     % \multirow{2}{*}{ $8\times$}   & w/o fre & 30.04+1.51 &0.748±0.034  & - \\
     % & Ours & 37.94±1.66 &0.975±0.006 & -  \\
    \bottomrule
    \end{tabular}
    \end{adjustbox}
\label{tab:ablation_spa}
\end{table}

\subsubsection{Analysis on Frequency Domain Fusion.}
To verify the effectiveness of the proposed selective frequency fusion module, we design the following experiments:  (1) fusion through element-wise addition and (2) fusion through our proposed SFF module, denoted as `Sum' and `SFF' in Table \ref{tab:ablation_fre}, respectively.
The results presented in the table show that element-wise addition of features from two modalities results in suboptimal performance. In contrast, our proposed SFF module significantly enhances performance, achieving a 0.43 dB improvement in PSNR under 4$\times$ acceleration and a 0.21 dB improvement under 8$\times$ acceleration.

\section{Conclusion}
\label{Con}
This study explores the comprehensive and efficient integration of complementary information across modalities for multi-modal MRI reconstruction. 
We present the MMR-Mamba framework, which effectively integrates information through the TCM module in the spatial domain and the SFF module in the frequency domain, along with integrating the spatial-frequency features through the ASFF module.
In particular, the TCM module employs cross Mamba blocks to selectively supplement complementary information from the reference modality to the target modality, while the SFF module integrates global information in the Fourier domain and restores high-frequency signals for reconstructing structural details.
We conducted extensive experiments on the BraTS and fastMRI knee datasets, demonstrating the superiority of our proposed framework in reconstructing MRI under different acceleration factors.
%
% This work provides valuable insights for advancing research in multi-modal MRI reconstruction, particularly in optimizing information fusion across spatial and frequency domains.

\begin{table}[t]
    \centering
    % \footnotesize
    \caption{
    \textbf{Ablation Study on Frequency Domain Fusion.} We report mean±std for the PSNR, SSIM, and NMSE metrics on the BraTS dataset under $4\times$ and $8\times$ acceleration. 
    }
    % \scriptsize
    \begin{tabular}{c|c|ccc}
    \toprule
    % \rowcolor[HTML]{EFEFEF}  \multirow{2}{*}{IXI}
     % & \textbf{BraTS}  &\multicolumn{3}{c}{$4 \times$} \\
     % \midrule
     & Method & PSNR$\uparrow$ & SSIM$\uparrow$ & NMSE~$(10^{-2})$$\downarrow$  \\
     \midrule
     \multirow{2}{*}{ $4\times$}   & Sum & 40.06±1.79 &0.982±0.005  & 0.560±0.002 \\
     & SFF & \textbf{40.49±1.84} & \textbf{0.984±0.005} & \textbf{0.509±0.002} \\
     \midrule
     \multirow{2}{*}{ $8\times$}   & Sum & 37.01+1.77 & 0.970±0.008 & 1.128±0.004 \\
     & SFF & \textbf{37.22±1.79} & \textbf{0.971±0.008} & \textbf{1.078±0.004}  \\
     % \multirow{2}{*}{ $8\times$}   & w/o fre & 30.04+1.51 &0.748±0.034  & - \\
     % & Ours & 37.94±1.66 &0.975±0.006 & -  \\
     \bottomrule
    \end{tabular}
\label{tab:ablation_fre}
\end{table}

\bibliographystyle{IEEEtran}
\bibliography{IEEEabrv,Reference}
\end{document}